\documentclass[aps,11pt,draft,notitlepage,oneside,onecolumn,nobibnotes,nofootinbib,preprintnumbers,superscriptaddress,showpacs,centertags,showkeys,amsmath,amssymb]{revtex4}

\usepackage{graphicx} 
\usepackage{dcolumn} 

\def\phi{\varphi}

\begin{document}

\title{GENERATING FUNCTIONAL FOR BOUND STATES IN QED\footnote{Talk
(M. Pi\c{a}tek) given at the HADRON STRUCTURE '07
International Conference, Modra-Harm\`{o}nia, Slovakia, September
3--7, 2007.}}

\author{M. Pi\c{a}tek}
\email{piatek@theor.jinr.ru, piatek@ift.uni.wroc.pl,
radekpiatek@gmail.com}

\author{V.N. Pervushin}
\email{pervush@theor.jinr.ru}

\author{A.B. Arbuzov}
\email{arbuzov@theor.jinr.ru} \affiliation{Bogoliubov Laboratory
of Theoretical Physics, Joint Institute for Nuclear Research,
141980, Dubna, Russia}


\begin{abstract}
The manifestly Lorentz covariant formulation of quantum
electrodynamics disregards Coulomb instantaneous interaction and
its consequence --- instantaneous bound states (IBS's). In this
article we consider the way of construction the IBS generating
functional using the operator generalization of the initial data
in the Dirac Hamiltonian approach to QED.
\end{abstract}



\maketitle

\section{Introduction}
The first papers by Dirac \cite{Dirac}, Heisenberg, Pauli
\cite{hp}, and Fermi \cite{Fermi} on the quantization of
electrodynamics ran into difficulties of the determination of
physical variables. The interpretation of all gauge components as
independent variables contradicts the quantum principles, whereas
excluding nonphysical variables contradicts  the relativistic
principles.

The first quantization of electrodynamics belongs to Dirac
\cite{Dirac} who disregarded the relativistic principles and
excluded nonphysical components  by the reduction of the initial
action to the  solution of the Gauss law constraint, i.e. the
equation for the time-like component of the gauge vector field.
The Gauss law connects initial data of the time-like component
with the data of all other fields. But this  {\it
constraint-shell} method had a set of defects, including
nonlocality, explicit noncovariance as the dependence on the
external time axis of quantization, and complexity.
Feynman-Schwinger-Tomonaga formulation of QED admitted a simpler
method based on the  extended dynamics where all components were
considered on  equal footing with fixing a relativistic gauge.

At the beginning of the sixties of the twentieth century, Feynman
found that the naive generalization of his method of  construction
of QED did not work for the non-Abelian theory. The unitary
S-matrix in the non-Abelian theory was obtained in the form of the
Faddeev-Popov (FP) path integral \cite{fp} by the brilliant
application of the theory of connections in vector bundle. There
is an opinion that the FP path integral is the highest level of
quantum description of gauge relativistic constrained systems.  In
any case, just this FP integral was the basis to prove
renormalizability of the unified theory of electroweak
interactions in papers by 't Hooft  and Veltman warded the Nobel
prize in 1999.

Nevertheless, in the context of the first Dirac quantization and
its Hamiltonian generalizations~\cite{Dirac2}\cite{gpk}, the
intuitive status of the FP integral was so evident that  two years
after the paper \cite{fp}, Faddeev gave the foundation of the FP
integral by the construction of the unitary S-matrix
\cite{faddeev69} for an ``equivalent non-Abelian unconstrained
system'' derived by resolving constraints in terms of the {\it
radiation variables}  of the Hamiltonian description.

Faddeev  showed that on the one hand, the constraint-shell
dynamics is compatible with the simplest quantization by the
standard Feynman path integral, on the other hand, this Feynman
integral is equivalent to the FP integral in an arbitrary gauge.
This equivalence was proved by the change of variables in the
Feynman path integral that removed the time-like vector of the
canonical quantization into the phase factors of physical source
terms. These phase factors disappear for  S-matrix elements on the
mass shell of elementary particles. In other words, Faddeev proved
the equivalence of the constraint-shell approach with quantization
of gauge theories by the gauge-fixing method only for scattering
amplitudes \cite{faddeev69} where all color particle-like
excitations of the fields are on their mass-shell. But the
scattering amplitudes for color particles are nonobservable in
QCD. The  observables are hadrons as colorless bound states where
elementary particles are off mass-shell. Just for this case, the
Faddeev  theorem of equivalence of different gauges becomes
problematic even for QED in the sector of instantaneous bound
states, as the FP integral in a relativistic gauge loses all
propagators with  analytic properties that lead to instantaneous
bound states identified with observable atoms.

It is unknown how to generalize Faddeev equivalence theorem onto
bound states even for quantum electrodynamics. In this article we
present a construction of a generating functional for bound states
in QED using another possibility. We generalize the Dirac
constraint-shell approach to gauge theories. It seems, that this
method can lead to the relativistic covariant unitary
renormalizable scattering theory of instantaneous bound states in
QED. An extension of this framework to non-Abelian gauge theories
paves a way for a description of  hadronization and confinement in
QCD. This is our main motivation for this line of research.

\section{Bound State S-matrix in QED: statement of the problem}
To state the problem let us recall the Dirac approach \cite{Dirac}
to QED and its relation to the standard Faddeev-Popov \cite{fp}
path integral quantization of gauge theories. Our starting point
is the well known QED action:
\begin{equation}
\label{QED} W[A,\psi,\bar \psi]=\int
\textrm{d}^4x\bigg\{-\frac{1}{4}\left[\partial^\mu A^\nu
\!-\!\partial^\nu A^\mu\right]^2\!+\! \bar \psi
[i\gamma^{\mu}\partial_{\mu}\!-\! m]\psi\!+\! A_\mu
j^{\,\mu}\bigg\},
\end{equation}
where $A_{\mu}$ is a vector potential, $\psi$ is the Dirac
electron-positron bispinor field and $j_{\mu}=e  \bar {\psi}
\gamma_{\mu} \psi$ is the charge current. Dirac \cite{Dirac}
proposed to eliminate  the time component of the four-vector
potential by the substitution of the manifest resolution of the
Gauss constraint $\delta W/{\delta A_0}=0$ into the initial action
(\ref{QED}) in the rest frame $\ell^{(0)}_{\mu}=(1,0,0,0)$. A
result of such reduction procedure written in terms of the so
called radiation variables\footnote{Dirac's radiation variables
are gauge-invariant functionals of the initial fields $A_{\mu}$,
$\psi$, $\bar\psi$ describing physical degrees of freedom. How to
construct these observables, see \cite{Pervu,Polubarinov}.}
$A^{*}_{a}=(A_{1}^{*}, A_{2}^{*}), \psi^*, \bar\psi^*$ leads to
the following constraint-shell action:
\begin{eqnarray}
\label{DiracQED} W^{*}[A^*,\psi^*, {\bar
\psi}^*]\!&=&\!\! W\Big|_{\delta W/{\delta A_0}=0}\nonumber\\
&=&\int \textrm{d}^4x \frac{1}{2} \left[\sum_{a=1,2}(\partial_\mu
A_a^*
\partial^\mu A_a^*) \!+ \!\frac{1}{2}j_{0}^{*}\! \frac{1}{\Delta}j_{0}^{*}
\!-\! j_{k}^{*} A_{k}^{*} + {\bar {\psi}}^{*}
(i\gamma^{\mu}\partial_{\mu} \!-\! m){\psi}^{*}\!\right],\\[8pt]
&&\hspace{-122pt}\textrm{where}\;\;\;\;A_{k}^{*}=\sum\limits_{a=1,2}A^*_{a}e_{k}^{a},
\hspace{10pt} k=1,2,3;\hspace{35pt}\textrm{and}\hspace{35pt}
\frac{1}{\Delta}j_{0}^{*}(\,\mathbf{x},t)\stackrel{\rm
def}{=}-\frac{1}{4\pi}\int \textrm{d}^3 y
\frac{j_{0}^{*}(\,\mathbf{y},t)}{|\;\mathbf{x}-\mathbf{y}|}\nonumber.
\end{eqnarray}
Having (\ref{DiracQED}), one can define the constraint-shell path
integral:
\begin{equation}
\label{Zcs} Z^{*}[ s^*, {\bar {s}}^*, J^*|\ell^{(0)} ]\;=\;\int
\prod_{a=1,2} DA^*_a D\psi^*D{\bar
\psi}^*\textrm{e}^{iW^{*}[A^*,\psi^*, {\bar \psi}^*] + i S^* },
\end{equation}
where
\begin{equation}
S^*\;=\;\int \textrm{d}^4x \left({\bar s}^* \psi^* + {\bar \psi}^*
s^* +\sum\limits_{a=1,2}J^*_a A^*_a \right)
\end{equation}
is the external source term. The theory defined by (\ref{Zcs}) is
frame-depended. This frame dependence can be removed by two steps
\cite{faddeev69}. First, we make the identical change of radiation
variables by Lorenz-type ``dummy'' variables:
\begin{equation}
\label{newV} A_{k}^{*}[A^F]= A^F_{k}-\partial_{k}\Lambda{(A^F)},
\;\;\;\;\;\;\;\;\;\;\;\;\;\; \psi^{*}[A^F] =\exp\left\lbrace ie
\Lambda{(A^F)}\right\rbrace\psi,
\end{equation}
where $\Lambda{(A^F)}=\Delta^{-1}\,\partial^{j}A^F_{j}$. Then the
constraint-shell generating functional can be rewritten in the
following form
\begin{eqnarray}
Z^{*}[s^*, {\bar s}^*, J^*|\ell^{\,(0)}]\;&=&\;\int
\prod_{\mu}DA^F_\mu D \psi^F D{\bar \psi}^F \Delta _{FP}^F
\delta (F(A^F))\times\nonumber\\
&\times & \textrm{e}^{iW[ A^F,\psi^F, {\bar \psi}^F ] +
S^*[\ell^{(0)}\,]}.
\end{eqnarray}
The change of variables (\ref{newV}) introduces the additional
nonphysical degrees of freedom and the Faddeev-Popov determinant
of the transition to new variables of integration. The nonphysical
degrees of freedom are removed by delta function gauge condition.
However, the dependence on the time axis is still present in the
source term. To remove the frame dependence completely we make the
change of sources:
\begin{equation}\label{ch-1}
S^*[\,\ell\,] \Rrightarrow S^F, \;\;\;\;\;\;\;\;\;\;\; S^F=\int
\textrm{d}^4x \left( {\bar s}^F \psi^F + {\bar \psi}^F s^F +
A^F_{\mu}J^{\mu}\right).
\end{equation}
After that the constraint-shell generating functional $Z^{*}[ s^*,
\bar s^{*}, J^* |\ell^{(0)}]$ takes the equivalent form of the
Faddeev-Popov integral:
\begin{eqnarray}
\label{FP} Z^{F}[s^F, {\bar s}^F, J^F]&=& \int \prod_{\mu}DA^F_\mu
D \psi^F D{\bar \psi}^F \Delta _{FP}^F \delta (F(A^F))\times\nonumber\\
&\times& \textrm{e}^{iW[ A^F,\psi^F, {\bar \psi}^F ] + S^F}.
\end{eqnarray}

Thus the fundamental constraint-shell generating functional
(\ref{Zcs}) coincides with the Faddeev-Popov integral (\ref{FP}),
if the change of sources (\ref{ch-1}) is valid. However, the
validity of this change  was  shown only in the sector of
scattering amplitudes derived from Green functions for elementary
particles on their mass-shell \cite{faddeev69}.

On the other side it is unknown how to generalize the
Faddeev-Popov path integral approach onto instantaneous bound
states formed by the Coulomb singularity. Moreover nobody proved
that quantum electrodynamics based on $Z^F$ contains instantaneous
(observed) bound states. Really, the Faddeev-Popov perturbation
theory in the relativistic gauge (\ref{FP}) contains only photon
propagators with the light-cone singularities forming the
Wick-Cutkosky bound states \cite{Kummer} with the spectrum
different from the observed one. These bound states have the
problem of tachyons and the probability interpretation.

Therefore the question appears what is the Faddeev equivalence
theorem for instantaneous bound states, where elementary particles
are off mass-shell?

Our proposition is based on the generalization \cite{Pervu2} of
the described above formulation, where the parameters of the
comoving frame of reference in the Dirac approach to QED become
eigenvalues of the total momentum operator of instantaneous bound
states:
\begin{equation}
\ell_\mu^{(0)}\to\ell_\mu\to\hat \ell_\mu\sim \dfrac{\partial
}{\partial X^\mu}, \;\;\;\;\;\;\;\;\;\;
X^\mu=\sum\limits_{J=1}^{N} x^\mu_J,
\end{equation}
so that these eigenvalues are proportional to total momenta of
bound states
\begin{equation}
\hat
\ell^\mu\,\left|\Phi_B(P_B)\right\rangle=\dfrac{P^\mu_B}{M_B}\,\left|\Phi_B(P_B)\right\rangle,
\;\;\;\;\;\;\;\;\;\; P^\mu P_\mu=M_B^2.
\end{equation}
This framework yields the observed spectrum of bound states which
corresponds to the instantaneous Coulomb interaction 
and paves a way for constructing an instantaneous bound state
(IBS) generating functional:
\begin{equation}
Z^*[\,\ell^{(0)}\,]\to Z^*[\,\ell\,]\to Z^*_{\sf
IBS}[\,\hat\ell\,],
\end{equation}
and IBS S-matrix.

\section{Bound State Generating Functional in QED}
The instantaneous Coulomb interaction is described in
(\ref{DiracQED}) by the zero component current-current term, which
can be converted into the following form:
\begin{eqnarray}
\label{int} W_{\textsf{Coulomb}}^{*}[\ell^{(0)}]&=&\frac{1}{2}\int
\textrm{d}^{4}x_1 \textrm{d}^{4}x_2\,
\psi(x_2)\bar\psi(x_1){\cal K}(x_1,x_2|\ell^{(0)})\psi(x_1)\bar\psi(x_2)\nonumber\\
&\equiv&  \frac{1}{2}\left(\psi_2 \bar\psi_1, {\cal
K}_{(1,2)}(\ell^{(0)})\psi_1 \bar\psi_{2}\right),
\end{eqnarray}
where ${\cal K}(x_1,x_2|\ell^{(0)})=\ell^{
(0)\nu}\gamma_{\nu}V(|\,z_{\mu}^{\bot}(\ell^{(0)})|)\ell^{
(0)\nu}\gamma_{\nu} \delta(z_{\mu}\ell^{
(0)\mu})=\gamma_{0}V(|\,\mathbf{x}_1 -
\mathbf{x}_2|)\gamma_{0}\delta(x_{(1)0}-x_{(2)0})$ is the Coulomb
kernel written in the rest frame $\ell^{(0)}_{\mu}=(1,0,0,0)$. The
Coulomb kernel contains the Coulomb potential $
V(|\,z_{\mu}^{\bot}(\ell^{(0)})|)=-e^2\left[4\pi
|\,z_{\mu}^{\bot}(\ell^{(0)})|\right]^{-1}=
-e^2\left[4\pi|\,\mathbf{x}_1 - \mathbf{x}_2|\right]^{-1}$
depending on the transverse components
$z_{\mu}^{\bot}(\ell^{(0)})=z_{\mu}-\ell^{(0)}_{\mu}(z_{\nu}\ell^{(0)\nu})$
of the relative coordinate $z_{\mu}=x_{(1)\mu}-x_{(2)\mu}$. Since
the theory is Lorentz covariant \cite{Zumino,Polubarinov,Pervu},
the Coulomb kernel can be transformed to an arbitrary frame $
{\cal K}(x_1,x_2|\ell^{(0)}) \Rrightarrow {\cal K}(x_1,x_2|\ell)
=\ell^{\nu}\gamma_{\nu}V(|\,z_{\mu}^{\bot}(\ell)|)\ell^{\nu}\gamma_{\nu}
\delta(z_{\mu}\ell^{\mu}). $ Such form of ${\cal K}(x_1,x_2|\ell)$
can be generalized to the Coulomb kernel being an operator
function acting on the product of fermion fields $ {\cal
K}(x_1,x_2|\ell)\psi(x_1)\bar\psi(x_2)\Rrightarrow{\cal
K}(X,z|\hat\ell)\psi(z+X/2)\bar\psi(-z+{X}/{2})$, where
$\hat\ell_{\mu}\sim\frac{\partial}{\partial X_{\mu}}$ and
$X_{\mu}=\scriptsize\frac{1}{2}(x_1+x_2)_{\mu}$. The action of
this operator can be fully understood in the Fourier analysis.

The next step in our construction is a redefinition of the action
(\ref{int}) in terms of {\it bilocal fields} ${\cal
M}(x_1,x_2)={\cal M}(X,z)$ and generalized Coulomb kernel ${\cal
K}(X,z|\hat\ell)$ by means of the Legendre transformation
\cite{CKS,PRE}:
\begin{equation}
\label{Legendre} \frac{1}{2}\left(\psi_{2}\bar\psi_{1}, {\cal
K}_{(1,2)}(\hat\ell)\psi_{1}\bar\psi_{2}\right)
=-\frac{1}{2}\left({\cal M}, {\cal K}^{-1}(\hat\ell){\cal
M}\right)+\left(\psi_1\bar\psi_2, {\cal M}\right).
\end{equation}
Bilocal fields obey the Markov-Yukawa condition
\cite{MarkovYukawa}:
\begin{equation}
\label{MY} z_{\mu}\frac{\partial{\cal M}(X,z)}{\partial
X^{\mu}}=0.
\end{equation}
This constraint realized the physical definition of a bound
state\footnote{One of the first definitions of the physical bound
states in QED belongs to Lord Eddington \cite{Eddington}: ``A
proton yesterday and an electron today do not make an atom''. },
the so called ``simultaneity principle'', which states that we can
observe experimentally two particles as the bound state ${\cal
M}(x_1,x_2)$ at one and the same time:
$$
{\cal M}(x_1,x_2)={\cal M}(z,X)\Big|_{\sf
Atom~at~rest}\equiv\textrm{e}^{iM
X_{0}}\,\Psi(\mathbf{z})\,\delta(z_{0}).
$$
In addition the equation (\ref{MY}) has deep mathematical meaning
\cite{Kalinovsky,LukierskiOziewicz} as the constraint of
irreducible nonlocal representations of the Poincar\`e group for
an arbitrary bilocal field. Therefore the relativistic covariant
unitary perturbation theory in terms of such relativistic  bound
states can be constructed \cite{Kalinovsky}.

The Markov-Yukawa condition (\ref{MY}) is equivalent to our choice
of a reference frame in bilocal dynamics as a vector operator with
eigenvalues proportional to the total momenta of bound states
\cite{MarkovYukawa}. This is a key point of our construction, it
means that the constraint-shell QED allows us to construct the
bound state relativistic covariant perturbation theory. The most
straightforward way for constructing QED of instantaneous bound
states is to take the free constraint-shell QED action with the
generalized Coulomb instantaneous interaction\footnote{This
prescription neglects the ``retardation'' interaction. However,
one may argue, that at the point of the existence of the bound
state with the definite total momentum any instantaneous
interaction (\ref{int}) with the time axis parallel to this
momentum is much greater that any ``retardation'' interaction
\cite{love}.}(\ref{Legendre}). Then the instantaneous bound state
generating functional $Z^*_{\textsf{IBS}}[\,\hat \ell\,]$ can be
defined, which after an integration over fermions has the
following form:
\begin{equation}
\label{IBSgf} Z^{*}_{\sf IBS}[s^*, {\bar s}^*, J^*|\hat\ell\,]=
\langle\,*\,|\int{\cal D}{\cal
M}\textrm{e}^{iW_{\textsf{eff}}[{\cal M}]+iS_{\textsf{eff}}[{\cal
M}]} |\,*\,\rangle,
\end{equation}
where
\begin{eqnarray}
\label{eff} W_{\textsf{eff}}[{\cal M}]&=& \left(-i tr\log
(-G_{A}^{-1}+{\cal
M})\right)-\frac{1}{2}\left({\cal M}, {\cal K}^{-1}(\hat\ell){\cal M}\right),\\
S_{\textsf{eff}}[{\cal M}]&=& \left( s^* \bar s^*, (G_{A}^{-1}-{\cal M})^{-1}\right),\nonumber\\
-G_{A}^{-1}&\equiv
&\left(i\gamma^{\mu}\partial_{\mu}-ie\gamma^{\mu}A_{\mu}^{*}-m\right)\delta^{(4)}(x-y).\nonumber
\end{eqnarray}
The bracket $\langle\,*\,|\,\ldots\,|\,*\,\rangle$ means the
averaging over transverse photons:
$$
\langle\,*\,|\Upsilon|\,*\,\rangle=\int\prod\limits_{a=1,2} {\cal
D}A^{*}_{a}\textrm{e}^{iW[A^{*}]}\Upsilon.
$$

The effective field theory defined by the action
$W_{\textsf{eff}}[{\cal M}]$ reproduces the fermion and bound
state spectra. Indeed, let us determine the minimum of the
effective action (\ref{eff}):
$$
\frac{\delta W_{\textsf{eff}}}{\delta {\cal M}}\Big|_{{\cal
M}=\Sigma}=0,
$$
where $\Sigma$ is the corresponding classical solution for the
bilocal field. This stationarity equation can be rewritten in the
form of the Schwinger-Dyson equation \cite{Pervu}, which describes
the energy spectrum of Dirac particles in bound states. Then the
effective action can be expanded around the point of
minimum\footnote{The higher terms in the expansion (\ref{exp})
describe the interactions of bilocal fields. If the Coulomb
potential is replaced by any rising potential (QCD), these terms
are responsible for the spontaneous breakdown of chiral symmetry
\cite{Pervu2}.} ${\cal M}=\Sigma+\widetilde{{\cal M}}$,
\begin{equation}
\label{exp} W_{\textsf{eff}}(\Sigma+\widetilde{{\cal
M}})=W_{\textsf{eff}}^{(2)}+\ldots.
\end{equation}
The small fluctuations $\widetilde{{\cal M}}$ can be represented
as a sum over the complete set of orthonormalized solutions
$\Gamma$ of the equation:
$$
\frac{\delta^{2} W_{\textsf{eff}}}{\delta {\cal
M}^{2}}\Big|_{{\cal M}=\Sigma}\Gamma=0.
$$
This equation reproduces the Bethe-Salpeter equation
\cite{Salpeter}, which describes the spectrum of bound states (see
\cite{Pervu}).

\section{Summary and Open Problems}
The relativistic invariant generating functional for instantaneous
bound states (\ref{IBSgf}) and their amplitudes constructed in
this paper by means of the operator generalization of the initial
data in Dirac's radiation variables states the following problems:
1) building of a bound state S-matrix; 2) a proof of
renormalizability. The proof of renormalizability can be achieved,
because the main difference of IBS functional from the FP
functional for Lorentz gauge formulation is only the source term.
We hope that obtained IBS functional can be useful for studying
the physics of  the QED bound states like positronium, and it can
help us to clear up the QCD hadronization problems.

\section*{Acknowledgements}
M.P. would like to thank the Organizing Committee of the HADRON
STRUCTURE '07 International Conference for possibility to present
the report.

\end{document}